\documentstyle[prl,aps,epsf,epsfig,floats,twocolumn,amssymb]{revtex}
\newcommand{\cF}{{\cal F}}
\newcommand{\cL}{{\cal L}}
\newcommand{\cZ}{{\cal Z}}
\newcommand{\s}{\sigma}
\newcommand{\be}{\begin{equation}}
\newcommand{\ee}{\end{equation}}
\newcommand{\ba}{\begin{eqnarray}}
\newcommand{\ea}{\end{eqnarray}}
\newcommand{\vp}{{\bold{p}}}
\newcommand{\vq}{{\bold{q}}}
\newcommand{\vx}{{\bold{x}}}
\newcommand{\vy}{{\bold{y}}}
\newcommand{\tp}{\dot{\psi}}
\newcommand{\barr}[1]{\not\mathrel #1}
\begin{document}
\draft
\twocolumn[\hsize\textwidth\columnwidth\hsize\csname
@twocolumnfalse\endcsname

\title{
\hfill{\small { FZJ-IKP(TH)-2001-02}}\\[0.2cm]
Chiral Lagrangians at finite density}

\author{Jos\'e A. Oller}

\address{Forschungszentrum J\"ulich, Institut f\"ur Kernphysik (Th), D-52425 
 J\"ulich, Germany }
\maketitle


\begin{abstract}
The effective $SU(2)$ chiral Lagrangian with external sources is given in the 
presence of non-vanishing nucleon densities by calculating the in-medium contributions 
of the chiral pion-nucleon Lagrangian. As a by 
product, a relativistic quantum field theory for Fermi many-particle systems 
at zero temperature is directly derived from relativistic quantum field 
theory with functional methods.  
\end{abstract}
\medskip
{PACS numbers: 12.39.Fe, 24.85.+p, 13.75.L, 13.75.Gx}

] 

\vspace{1cm}

\noindent

\noindent
{\bf 1. }In the limit of massless up and down quarks the QCD Lagrangian is symmetric 
under the chiral group $SU(2)_L\times SU(2)_R$. One assumes that this 
symmetry is spontaneously broken to the diagonal subgroup $SU(2)_{L+R}$ giving rise 
to the appearance of 3 massless Goldstone bosons which finally acquire small masses 
due to the non-vanishing mass of the $u$ and $d$ quarks.

This symmetry breaking scenario constrains so much 
the interactions of the Goldstone bosons that the QCD Green functions can be 
calculated at low energies as an expansion in powers of momenta and quark 
masses. This is known as Chiral Perturbation Theory\cite{Wein,GL}.

 The extension 
of the theory to the case of low temperature at zero density was considered in
 ref.\cite{GLT}. In this article we study the case of small densities at zero 
temperature and derive 
 the corresponding chiral Lagrangian by calculating the in-medium contributions 
due to the chiral pion-nucleon Lagrangian\cite{gasser,fettes} with functional 
methods. Although we focus our treatment to 
QCD, the 
relativistic many-body formalism here deduced for Fermi~ systems can be applied to 
processes governed by other dynamical theories, as the traditional non-relativistic 
zero temperature many-body \cite{morris,fet} quantum theory which stresses the 
diagrammatic approach. Compared with standard 
quantum field theory 
at finite temperature $T$\cite{matsu,martin} in the grand canonical ensemble, 
one avoids the use of unknown chemical potentials which themselves have to be 
calculated in terms of the many-body forces. The former is accomplished by following 
quantum field theory at $T=0$ considering directly the change of the ground state 
from the vacuum to one with finite fermionic densities. In the same way one also avoids 
the non-trivial $T\rightarrow 0$ limit due to the so called anomalous 
diagrams\cite{lutt,nege}. The price to pay is to rely on the adiabatic 
hypothesis in order to determine the interacting ground state from that of the 
free case by turning on the interactions adiabatically. 

\vskip 15pt
\noindent
{\bf 2. }Let us take first the case of symmetric and unpolarized nuclear matter, the 
extension of the formalism to the asymmetric and polarized case is straightforward 
and will be shown below.  In the following we take the Heisenberg picture and, 
following closely scattering theory\cite{weinb}, we consider two ground states 
$|\Omega_{\hbox{out}}\rangle$ and $|\Omega_{\hbox{in}}\rangle$ which, under the action 
of any time dependent operator {\it at asymptotic times} $t\rightarrow \pm \infty$, 
respectively, behave as  two symmetric Fermi seas of free protons and neutrons. 
The Fermi seas are filled up to the corresponding 
baryonic density, 
{\scriptsize $ \displaystyle{\prod^{N}}$}$a^\dagger(\vp_n)|0\rangle$,  where the 
label $n$ includes also the spin and isospin indices, $N$ is the number of 
momentum states inside the Fermi sea with Fermi momentum $k_F=(3 \pi^2 \rho/2)^{1/3}$,
 $\rho$ is the total nuclear density and $|0\rangle$ is the vacuum. Our objective is 
to evaluate the generating functional ${\mathcal{Z}}[v,a,s,p]$ in 
the presence of vector $v_\mu$, axial $a_\mu$, scalar $s$ 
and pseudoscalar $p$ external fields\cite{GL} by working out the transition 
amplitude $\langle \Omega_{\hbox{out}}|\Omega_{\hbox{in}}\rangle_J$, where 
the label $J$ just indicates the presence of the aforementioned external sources. 
In this way by taking functional 
derivatives of ${\mathcal{Z}}[v,a,s,p]$ with respect to the external sources one 
evaluates the in-medium QCD connected Green functions 
(space-time averages at finite density of the quark currents coupled to the $v$, 
$a$, $s$ and $p$ sources). To do this we consider the effective 
chiral 
Lagrangians $\cL=\cL_{\pi\pi}+\cL_{\bar{\psi}\psi}+
\cL_{\bar{\psi}\psi\bar{\psi}\psi}+...$ with increasing number of pairs of nucleon 
fields $\psi(x)$. We first restrict ourselves to the term with no nucleon fields 
$\cL_{\pi\pi}$ and to that containing two of them 
$\cL_{\bar{\psi}\psi}=\bar{\psi}(x)D(x) \psi(x)$, together with the previous external 
fields. We will discuss later a way to include perturbatively the contributions 
of Lagrangians with higher number of nucleons by 
considering them to arise  from bilinear vertices through the exchange of an 
arbitrary heavy particle. Indeed, although we are 
talking about CHPT, the only thing that 
matters for the following derivations is that $\cL_{\bar{\psi}\psi}$ is bilinear in the 
fermions. 
Consider now the transition 
amplitude for the ground states from $t\rightarrow -\infty$ to 
$t\rightarrow +\infty$ in the presence of the previous external sources together with 
Grassmann sources $\eta$ and $\eta^\dagger$, coupled to the nucleon fields:

\begin{eqnarray}
\label{tran}
&&\langle \Omega_{\hbox{out}}|\Omega_{\hbox{in}}\rangle_{J,\eta,\eta^\dagger} =
\!\int\! [d U][d \psi]
[d \psi^\dagger] \langle \Omega_{\hbox{out}}|\psi(+\infty)\rangle \times 
\nonumber \\ 
&e&^{i \!\int\! dx \left[ {\mathcal{L}}_{\pi\pi} + \bar{\psi}D\psi+\eta^\dagger\psi+
\psi^\dagger\eta \right]} \langle \psi(-\infty)|\Omega_{\hbox{in}}\rangle
\end{eqnarray}
with the pion fields described by the $2\times 2$ unitary matrix $U$.

The ground state functional $\langle \psi(\pm\infty)|\Omega_{{\tiny \begin{array}{c}
\hbox{out}\\\hbox{in}\end{array}}}\rangle$ can be expressed in terms 
of that of the vacuum  by writing the $a(\bold{p}_n)_{\!\!{\tiny \begin{array}{c}
\hbox{out}\\\hbox{in}\end{array}}}$ operators as a function of 
$\psi(x)$ and its time derivative $\dot{\psi}(x)$:
\ba
a(\vp_n)_{\!\!{\tiny \begin{array}{c}
\hbox{out}\\\hbox{in}\end{array}}}\!=\!\lim_{{\tiny t\rightarrow \pm \infty }}
\frac{E(p)}{2m_N}e^{iE(p)t}\!\int\! 
d\vx\,e^{-i\vp \vx}\bar{u}(\vp_n) [\psi_n(x)
\nonumber \\
+\frac{i}{E(p)}\tp_n(x)],
\ea 
where as usual in scattering theory for $t\rightarrow \pm\infty$ the matrix elements 
are calculated as if there were no interactions. In the former expression, $E(p)$ 
is the 
energy of the nucleon with three-momentum $\vp$, $m_N$ is the nucleon 
mass and $u(\vp_n)$ is a Dirac spinor. Expressing $\tp(x)$ in terms of $\psi(x)$ (one 
way is using the Dirac equation), taking into account that $\psi(x) 
e^{i\!\int\! dy\, \eta^\dagger(y) \psi(y)}=-i
\frac{\stackrel{\rightarrow}\delta}{\delta \eta^\dagger(x)}
e^{i\!\int\! dy\,\eta^\dagger(y)\psi(y)}$, the analogous expression for 
$\psi^\dagger(x)$, and substituting all that in eq.(\ref{tran}), one obtains for 
$\langle \Omega_{\hbox{out}}|\Omega_{\hbox{in}}\rangle_{J,\eta,\eta^\dagger}$:
\ba
\label{tran8}
&&\lim_{\tiny{\begin{array}{c}t\rightarrow +\infty \\ t'\rightarrow -\infty \end{array}}}
\!\int\! [d U] (\hbox{det}\,D)  e^{i\!\int\! dx \cL_{\pi\pi}}  \left(\prod_{n}^N
\frac{E(p_n)}{2m_N}\!\int\! d\vx_n\, 
e^{ip_n x_n} \right. \nonumber \\ && \left.\bar{u}(\vp_n) \bigg[ 1-  
\frac{i}{E(p_n)}(\gamma^0 \sum_{j=1}^3\gamma^j \frac{\partial}{\partial x_n^j}
+i \gamma^0 m_N )\bigg]\frac{\stackrel{\rightarrow}{\delta}}{\delta \eta^\dagger(x_n)}
\right) \nonumber \\ &&e^{-i\!\int\! dx \!\int\! dy\, \eta^\dagger(x)D^{-1}(x,y) 
\gamma^0 \eta(y)}\prod_{m}^N \frac{E(q_m)}{2m_N}\!\int\! d\vy_m\,e^{-iq_m y_m} 
\nonumber \\ && 
\Bigg[ 
\frac{\stackrel{\leftarrow}{\delta}}{\delta \eta(y_m)}-\frac{i}{E(q_m)} 
\bigg(\sum_{k=1}^3\frac{\partial}{\partial y_m^k} 
\frac{\stackrel{\leftarrow}{\delta}}{\delta \eta(y_m)}\gamma^k\gamma^0+i 
\frac{\stackrel{\leftarrow}{\delta}}{\delta \eta(y_m)} 
 \nonumber \\ && 
\gamma^0 m_N \bigg)\Bigg] \gamma^0\,u(\vq_m) 
\ea
where the integration over the nucleon fields and conjugate momenta is also done. 
Furthermore we define $x_n=(t,\vx_n)$, $y_m=(t',\vy_m)$, $p_n=(E(p_n),\vp_n)$ 
and analogously for $q_m$. The action of the 
spatial derivatives can be readily taken into 
account by integrating by parts. In this way, they only act on the corresponding 
exponentials giving rise to three-momenta factors that can be further simplified by 
applying the Dirac equation on the Dirac spinors. In this way we have:
\ba
\label{direq}
\bar{u}(\vp_n)\left[1-\sum_{j=1}^3\frac{p_n^j}{E(p_n)}\gamma^0\gamma_j+
\frac{m_N}{E(p_n)}\gamma^0\right]&=&\frac{2m_N}{E(p)} u^\dagger(\vp_n)\nonumber \\
\left[1-\sum_{j=1}^3\frac{p_n^j}{E(p_n)}\gamma^0\gamma_j+\frac{m_N}{E(p_n)}\gamma^0
\right]\gamma^0 u(\vp_n)&=&\frac{2m_N}{E(p)}u(\vp_n)\nonumber
\ea 
Applying the previous results to eq.(\ref{tran8}) it simplifies to:
\ba
\label{tranmore}
&&\lim_{\tiny{\begin{array}{c}t\rightarrow +\infty \\ t'\rightarrow -\infty \end{array}}}
\!\int\! [d U] (\hbox{det}\,D) e^{i\!\int\! dx \cL_{\pi\pi}}  \bigg(\prod_{n}^N
\!\int\! d\vx_n\, e^{i p_n x_n} 
u^\dagger(\vp_n)\nonumber \\&&
 \frac{\stackrel{\rightarrow}{\delta}}{\delta \eta^\dagger(x_n)}\bigg) 
\,e^{-i\!\int\! dx \!\int\! dy\,\eta^\dagger(x)D^{-1}(x,y) \gamma^0 \eta(y)}\,
\prod_m^N \!\int\! d\vy_m e^{-i q_m y_m}\nonumber \\&&
\frac{\stackrel{\leftarrow}{\delta}}{\delta \eta(y_m)} u(\vq_m) \nonumber
\ea
After acting with the left derivatives 
$\stackrel{\rightarrow}{\delta}\!\!/ \delta \eta^\dagger(x_n)$ on the 
exponential depending on the Grassmann sources, the former expression can be 
recast as:
\ba
\label{tran9}
&&\lim_{\tiny{\begin{array}{c}t\rightarrow +\infty \\ t'\rightarrow -\infty \end{array}}}
\!\int\! [d U] (\hbox{det}\,D) e^{i\!\int\! dx \cL_{\pi\pi}}  \bigg(\prod_{n}^N
\!\int\! d\vx_n\, e^{i p_n x_n}  \nonumber \!\int\! dz_{n'} \\ && 
u^\dagger(\vp_n)  D^{-1}(x_n,z_{n'}) \gamma^0 \eta(z_{n'}) \bigg) 
e^{-i\!\int\! dx \!\int\! dy\,\eta^\dagger(x)D^{-1}(x,y) \gamma^0 \eta(y)}
\nonumber \\ && \prod_m^N \!\int\! d\vy_m e^{-i q_m y_m}
\frac{\stackrel{\leftarrow}{\delta}}{\delta \eta(y_m)} u(\vq_m). \nonumber
\ea
This result is equal to $e^{i\cZ[v,a,s,p]}$ when $\eta \rightarrow 0$ 
and $\eta^\dagger \rightarrow 0$.\footnote{Without taking the limit 
$\eta,\,\eta^\dagger \rightarrow 0$, the generating functional contains baryonic 
sources and taking differential derivatives with respect to them one could directly 
evaluate in-medium  baryonic Green functions, e.g. nucleon propagators. Nevertheless, 
since for our present purposes the baryons in the medium constitute just a 
background,  we do not consider this case any further.} Hence we can equalize to 
1 the second exponential from the left and the 
remaining right 
derivatives and $\eta(z_n)$ sources have to be paired in order to finish with a 
non-vanishing result. Thus we have:
\ba
\label{Z}
&&e^{i\cZ}=\lim_{\tiny{\begin{array}{c}t\rightarrow +\infty \\ t'\rightarrow -\infty 
\end{array}}}\!\int\! [d U] (\hbox{det}\,D) e^{i\!\int\! dx \cL_{\pi\pi}}
 \sum_{\s}\epsilon(\s) \prod_{n}^N \!\int\! d\vx_n  \nonumber \\ && 
 \!\int\!  d\vx_{\s_n}
\,e^{i p_n x_n} u^\dagger(\vp_n) D^{-1}(x_n,x_{\s_n})\gamma^0 e^{-i p_{\s_n} x_{\s_n}}
 u(\vp_{\s_n}) 
\ea
with $\epsilon(\sigma)$ the signature of the permutation $\sigma$ over all the indices 
--momenta, spin and isospin. In order to continue let us write the operator $D(x)\equiv
D_0(x)-A(x)$ with $D_0(x)=i\gamma^\mu \partial_\mu-m_N$ the Dirac operator 
for the free motion of the nucleons. On the other hand, the operator $A(x)$ is 
completely general although in our case at hand, CHPT, it is subject 
to a chiral expansion of powers of soft three-momenta and quark masses. Furthermore, 
let us note that:
\ba
\label{D0}
\lim_{t \rightarrow +\infty}\!\int\! d\vx\, e^{i p x}u^\dagger(\vp)D_0^{-1}(x,x')=
-i\, e^{i p x'}\bar{u}(\vp) \nonumber \\
\lim_{t \rightarrow -\infty}\!\int\! d\vx\, D^{-1}_0(x',x)  \gamma^0 e^{-i p x} u(\vp)=
-i\, e^{-i px'} u(\vp)
\ea
This result can be easily obtained by writing $D_0^{-1}(x,x')$ in 
four-momentum space and then performing the integral over the temporal component of 
the momentum taking care of the imposed limits. For instance, let us take the first 
of the previous equations. Then we have:
\ba
\label{d0b}
&&\lim_{t \rightarrow +\infty}\!\int\! d\vx\, e^{i p x}u^\dagger(\vp)D_0^{-1}(x,x')=\nonumber 
\\&&
\lim_{t \rightarrow +\infty}\!\int\! d\vx\, e^{i p x}u^\dagger(\vp)
\int\!\! \frac{dR}{(2\pi)^4} 
\frac{e^{-i R(x-x')}(\barr{R}+m_N)}{R^2-m_N^2+i\epsilon},\nonumber
\ea
with $\epsilon$ a positive infinitesimal. Exchanging the order of the integrations, 
the spatial one gives rise to 
$(2\pi)^3 \delta({\bold R}-\vp)$ which fixes ${\bold R}$. As a result one has:
\ba
\label{int}
\lim_{t \rightarrow +\infty}\!\int\! \frac{dR^0}{2\pi} u^\dagger(\vp) 
\frac{e^{i p^0 t} e^{-i R^0(t-x'_0)}e^{-i\vp \vx'}}{R_0^2-\vp^2-m_N^2+i\epsilon}
\!\!\left(R^0\gamma^0\!-\! \vp {\bold \gamma}+m_N\right)\nonumber
\ea
Since $t\rightarrow +\infty$ then $t-x'_0>0$ and we close the integration contour over 
$R^0$ with a semi-circle of infinite radius on the lower half-plane picking up 
the pole at $R^0=p^0-i\epsilon=E(p)-i\epsilon$. Applying the 
Dirac equation to the result one  arrives to eq.(\ref{D0}). 

Then taking into account eq.(\ref{D0}) and 
the expansion $D^{-1}\gamma^0=[D_0-A]^{-1}\gamma^0=D_0^{-1}\gamma^0+
D_0^{-1}AD_0^{-1}\gamma^0+...$, we can rewrite eq.(\ref{Z}) as:
\ba
\label{Z2}
&&e^{i\cZ}=\!\int\! [dU](\hbox{det}\, D)\,e^{i\!\int\! dx \cL_{\pi\pi}}\sum_{\s}\epsilon(\s)
\prod_{n}^N \!\int\! d\vx_n \!\int\! d\vx_{\s_n} \nonumber \\ &&  
e^{-i\vp_n \vx_n}u^\dagger(\vp_n)\Bigg[ \delta(\vx_n-\vx_{\s_n})\delta_{n,\s_n}-
i\!\int\! dt \!\int\! dt'\,e^{i E(p_n) t}\gamma^0 
\nonumber \\ && 
A(x_n,x_{\s_n})\, e^{-i E(p_{\s_n}) t'} 
-i\!\int\! dt \!\int\! dt'\! \int\! dz\, dz'\,
e^{i E(p_n) t} \gamma^0 
 \nonumber \\ && 
 A(x_n,z) D_0^{-1}(z,z') A(z',x_{\s_n})e^{-i E(p_{\s_n}) t'}+...\Bigg]
\nonumber \\ &&
\, e^{i\vp_{\s_n}\vx_{\s_n}}u(\vp_{\s_n}) 
\ea
where now both $t$ and $t'$ are integration variables being the time 
components of 
$x_n$ and $x_{\s_n}$, respectively. The dots just refer to 
those terms with an increasing number of insertions of the operator $A$ coming from 
the geometric expansion of $D^{-1}=\left[D_0-A\right]^{-1}$ discussed above. 
Furthermore $A(x,y)$ is defined such that 
\ba
\int dx\,\bar{\psi}(x)A(x)\psi(x)\!=\!\int\!dx\, dy\,\bar{\psi}(x)A(x,y)\psi(y).
\nonumber
\ea 

As a result of eq.(\ref{Z2}) we can simply state that:
\be
\label{ez}
e^{i\cZ[v,a,s,p]}=\!\int\! [dU] (\hbox{det}\,D)\, e^{i\!\int\! dx \cL_{\pi\pi}}
\,(\widetilde{\hbox{det}}\,{\cal{F}})
\ee
where the tilde in $\widetilde{\hbox{det}}\,{\cF}$ indicates that the determinant 
has to be 
taken in the subspace of the Fermi sea states expanded by the basis functions 
$e^{i\vp_n \vx} u(\vp_n)$ with $|{\bold{p}}_n|<k_F$. In this notation $\cF$ is given by:
\ba
\label{F}
\cF\equiv I_3-i\!\int\! dt\!\int\! dt'\, e^{i H_0 t}\,\gamma^0 A\left[I_4-D^{-1}_0 A\right]^{-1}
\,e^{-i H_0 t'} 
\nonumber
\ea
where $I_3\equiv \delta(\vx_n-\vx_m)\delta_{n,m}$ and 
analogously $I_4\equiv \delta(x_n-x_m)\delta_{n,m}$. On the other hand, $e^{-i H_0 t'} 
e^{i\vp_n \vx'} u(\vp_n)=e^{-i p_n x'} u(\vp_n)$ and $e^{-i \vp_n \vx}u^\dagger(\vp_n) 
e^{i H_0 t}=e^{i p_n x} u^\dagger(\vp_n)$.

In order to obtain from eq.(\ref{ez}) the contributions of the surrounding medium to the 
generating functional it is convenient to 
exponentiate $\widetilde{\hbox{det}}\,\cF$ as $\exp(\widetilde{\hbox{Tr}}\log \cF)$ 
(where the tilde has the same meaning as before). Then we have:
\ba
\label{fd}
&&e^{i \cZ[v,a,s,p]}=\!\int\! [dU](\hbox{det}\,D) \exp\bigg\{i\!\int\! dx \cL_{\pi\pi} 
+\sum_{r=1}^2\sum_{\alpha=1}^2
\nonumber \\ && 
\!\int^{k_F}\hspace{-0.5cm} \frac{d\vp}{(2\pi)^3 2E(p)} \!\int\! d\vx \!\int\! d\vy\, 
e^{-i\vp \vx} u^\dagger_{r\alpha}(\vp)\,\log[\cF]
\arrowvert_{(\vx\,\alpha,\vy\,\alpha)}
 \nonumber \\ && 
\,e^{i\vp \vy} u_{r\alpha}(\vp) \bigg\}
\ea
where we have indicated explicitly the spin $r$ and isospin $\alpha$ indices. It is very 
appropriate to stress at this point that eq.(\ref{fd}), although formal, is 
non-perturbative. From 
this result we can readily read out the new effective chiral Lagrangian density 
$\tilde{\cL}_{\pi\pi}$ in the presence of a nuclear density just by equating the 
expression between curly brackets to $i\!\int\! dx\,\tilde{\cL}_{\pi\pi}$. 

\vskip 15pt
\noindent
{\bf 3. }The perturbative theory is obtained by expanding  $\log\cF$ in eq.(\ref{fd}), 
so that $e^{i{\cZ}}$ can be written as:
\ba
\label{expF}
&&\!\int\! [dU](\hbox{det}\,D) \exp\Bigg\{i\!\int\! dx\, \cL_{\pi\pi} -i\sum_{r=1}^2
\sum_{\alpha=1}^2 \!\int^{k_F}\hspace{-0.5cm} \frac{d\vp}{(2\pi)^3 2E(p)}
 \nonumber \\ &&   
 \!\int\! dx \, dy\, e^{ipx}\bar{u}_{r\alpha}(\vp)\,
A[I_4-D^{-1}_0 A]^{-1}\arrowvert_{(x\alpha,y\alpha)}\,e^{-i p y}u_{r\alpha}(\vp)
\nonumber \\ &&  
 +\frac{1}{2} \sum_{r,s=1}^2\sum_{\alpha,\beta=1}^2
\!\int^{k_F}\hspace{-0.5cm}\frac{d\vp}{(2\pi)^3 2 E(p)}\!\int^{k_F}\hspace{-0.5cm}\frac{d\vq}{(2\pi)^3 2E(q)}
\!\int\! dx\,dx'\,dy' \,dy
 \nonumber \\ &&  
 e^{i p x}\bar{u}_{r\alpha}(\vp)\,A[I_4-D_0^{-1}A]^{-1}\arrowvert_{(x\alpha,x'\beta)}
\,e^{-i q x'}u_{s\beta}(\vq)\, e^{i q y'} \bar{u}_{s\beta}(\vq)
 \nonumber \\ && 
A[I_4-D_0^{-1}A]^{-1}
\arrowvert_{(y'\beta,y\alpha)}\,e^{-i p y}u_{r\alpha}(\vp)+...\Bigg\}.\nonumber
\ea  
Finally, taking into account the relation 
{\scriptsize $\displaystyle{\sum_{r=1}^2}$}$ u_r(\vp)
\otimes \bar{u}_r(\vp)=\barr{p}+m_N$ we arrive at the following expression for the 
generating functional:
\ba
\label{fZ}
&&e^{i\cZ}=\!\int\! [dU](\hbox{det}\,D) \exp\Bigg\{i\!\int\! dx\, \cL_{\pi\pi}-i
\!\int^{k_F}\hspace{-0.5cm}\frac{d\vp}{(2\pi)^3 2 E(p)} 
\nonumber \\&&
\!\int\! dx\,dy\, e^{ip(x-y)}\,\hbox{Tr}\Bigg( A[I_4-D_0^{-1}A]^{-1}\arrowvert_{(x,y)}
(\barr{p+m_N})\Bigg)+
\nonumber \\ &&
\frac{1}{2}\!\int^{k_F}\hspace{-0.5cm}\frac{d\vp}{(2\pi)^32E(p)}\!\int^{k_F}
\hspace{-0.5cm}\frac{d\vq}{(2\pi)^3 2 E(q)}
\!\int\! dx\,dx'\,dy\,dy'\,e^{ip(x-y)}
\nonumber \\ &&
e^{-iq(x'-y')}\,\hbox{Tr}\Bigg(
A[I_4-D_0^{-1}A]^{-1}\arrowvert_{(x,x')}(\barr{q}+m_N)A[I_4-
\nonumber \\ &&
D_0^{-1}A]^{-1}\arrowvert_{(y',y)}(\barr{p}+m_N)\Bigg)+...\Bigg\}
\ea
where the trace refers both to the isospin and spinor indices. 
The previous formula implies a double expansion for obtaining the contributions 
of the nuclear medium to the chiral Lagrangian. One is the standard 
chiral expansion by expanding the {\it vacuum} operator 
$A[I_4-D_0^{-1}A]^{-1}=A+A D_0^{-1} 
A+ A D_0^{-1} A D_0^{-1}A+...$ together with $A$ itself, 
$A=A^{(1)}+A^{(2)}+A^{(3)}+...$ in increasing powers of  momenta and quark masses 
valid at low energies. Explicit expressions of $A$ up to ${\mathcal{O}}(p^3)$ can be 
obtained from ref.\cite{fettes}. The another is an expansion in the number of 
insertions of {\it on-shell} fermions 
belonging to the Fermi sea, schematically indicated in fig.1 by a thick solid line. 
Both expansions can be related by giving a chiral power couting to 
$k_F$ which can be naturally counted as ${\mathcal O}(p)$\cite{med2} since 
for nuclear saturation density $k_F\simeq 2 M_\pi$ with $M_\pi$ the pion mass. 
Moreover, the circles labeled by $\Gamma$ correspond to the non-local 
operator $-i A[I_4-D_0^{-1}A]^{-1}$. Note that when inserting $n$ Fermi seas from the 
expansion of the logarithm one picks up a factor $-(-1)^{n}/n$ where the global minus 
sign appears due to the fermionic closed loop, $n$ is a combinatoric 
factor because any cyclic permutation in the trace of $n$ Fermi seas with their 
associated $n$ $\Gamma$ operators gives the same result and finally the sign 
$(-1)^n$ is a pure in-medium factor that one has to keep in mind and is already 
present in standard many-body theory\cite{fet}.

\begin{figure}[htb]
\centerline{\epsfig{file=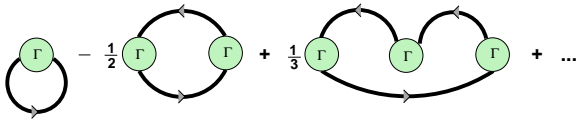,width=3.0in}}
\vspace{0.3cm}

\caption[pilf]{\protect \small
Diagrammatic expansion of eq.(\ref{fZ}). Every thick line corresponds to the insertion 
of a Fermi-sea and each circle to the insertion of an operator 
$\Gamma\equiv -i A\left[I_4-D_0^{-1}A\right]^{-1}$.}
\end{figure}

A generally non-local vertex $\Gamma$  comes from the iteration 
of the $A$ operator with intermmediate free baryon propagators $D_0^{-1}$, obeying 
the usual Feynman rules, see fig.2. Notice that $A$ was defined from the Lagrangain 
$\bar{\psi} D \psi$ removing the free term $\bar{\psi}D_0 \psi$ and changing the sign 
to the rest, this is why a minus sign appears in front of $A$ in fig.2. On the other 
hand $i\,D_0^{-1}(x,y)=i\,\int \frac{d^4p}{(2\pi)^4}
\frac{\barr{p}+m_N}{p^2-m_N^2+i\epsilon}$ is the usual 
baryon propagator.

\begin{figure}[htb]
\centerline{\epsfig{file=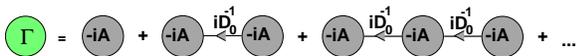,width=3.0in}}
\vspace{0.3cm}

\caption[pilf]{\protect \small
Expansion of the generalized non-local vacuum vertex $\Gamma$. Every solid line 
corresponds to a vacuum baryon propagator and each circle to the insertion of an
 operator $-i A$ from $\bar{\psi} D\psi$.}
\end{figure}

Hence a final diagram, when expanding $\Gamma$ up to the required accuracy, will be 
a set of $n\geq 1$ Fermi-sea insertions, 
$m\geq 0$ free baryon-propagators and of $m+n$ vertices $-i A$. First we include 
the  $-1$ global sign  because of the fermionic closed loop, and the 
combinatoric factor $1/n$  together with the sign $(-1)^n$. Then, following the 
diagram in the opposite sense to that of the fermionic arrows, write for 
each Fermi-sea an integral $\int^{k_F}\!\!\frac{d{\bold{p}}\,
(\barr{p}+m)}{(2\pi)^32E(p)}$ with $p^0=E(\bold{p})$, for each 
vaccum baryon propagator with free momentum $p$ write $ i 
\int\!\!\frac{d p}{(2\pi)^4}\frac{\barr{p}+m_N}{p^2-m_N^2+i\epsilon}$ and for a 
vertex in momentum space a term $-iA (2\pi)^4$, keeping in mind the energy-momentum 
conservation at each vertex. Finally sum over the spin and isospin indices of the 
fermions. This defines explicitely the Feyman rules in momentum 
space in order to obtain $i (2\pi)^4$ times the desired connected graph accompied by
 the global delta of energy-momentum conservation. Analogous Feynman rules hold of 
course in configuration space, e.g., see eq.(\ref{fZ}). 

An equivalent way to state the previous rules, without including explicitely the  
integral symbols and factors $2\pi$, is to write the same 
sign-combinatoric factor $-(-1)^n/n$ and vertices $-i A$ as before. Then for the free
 nucleons one has just the free propagator $i\frac{\barr{p}+m}{p^2-m^2+i\epsilon}$ and 
for the Fermi-sea baryons the factor $(2\pi)\delta(p^2-m^2)\theta(p^0)(\barr{p}+m)
\theta(k_F-|\bold{p}|)$. Finally sum over all the discrete indices attached to the 
fermions and integrate over all 
the free four-momenta with the measure $\int d^4 p/(2\pi)^4$ after taking into account 
energy-momentum conservation at each vertex.

Fig.1 fixes the skeleton structure of standard in-medium {\it vertices} since 
still one has to 
consider the pion fields contained in $A$ over which one has 
to integrate in eqs.(\ref{fd}), (\ref{fZ}) to finally obtain the generating 
functional. That is, from the vertices $A$ as well as from $\cL_{\pi\pi}$, one can 
generate internal as well as external (coupled to the sources) pionic legs denoted 
by dashed lines in figs.3a and 3b. Here one has essentially the same 
Feynamn rules than in vaccum in order to proceed in a perturbative way, simply 
for each pionic line with four-momentum $q$, one writes the vacuum propagator 
$i \int \frac{d q}{(2\pi)^4}\frac{1}{q^2-M_\pi^2+i\epsilon}$, for the first version 
of the in-medium Feynman rules. For the second one has 
$\frac{i}{q^2-M_\pi^2+i\epsilon}$. The important remark 
to keep in mind is that a generalized vertex has analogous properties to those 
of a  standard 
{\it local} quantum field theory one to the effects of determing the numerical factors 
accompanying the exchange of pion lines inside a given diagram. This can be seen just 
by applying 
standard perturbative 
techniques in path integrals to the action given between brackets in eq.(\ref{fd}) or 
more explicitely in eq.(\ref{fZ}). Several examples are discussed in detail in 
ref.\cite{med2}.  

\vskip 15pt
\noindent
{\bf 4. }Notice also that 
ultra-violet parts in the integration over running pionic momenta 
generate local multi-nucleon vertices and hence a proper treatment of pion loops 
can only be done by including simultaneously local nucleon interactions in order to 
reabsorb the divergences.  This is denoted by the squares in figs. 3c and 3d. 
It is important to realize that while in fig.3a the pion is exchanged inside the 
same vertex (in the generalized sense of above) in fig.3b the pions are 
exchanged between two of them. This implies the presence of only one 
Dirac trace in fig.3c (the flow of the Dirac indices along the 
propagators on both sides of the square is indicated by the arrows of the open 
solid lines) and  
two Dirac traces in fig.3d. Thus, when writing the matrix elements corresponding 
to diagrams of the type of figs.3c and 3d (with local interactions) one has 
to keep track of the number of closed traces in the spinor indices since each of 
them will lead to a sign $(-1)$ and its own combinatoric factor $1/n_j$ (with 
$n_j$ the number of Fermi-sea insertions in the closed Dirac loop) 
together with the sign $(-1)^{n_j}$. One can also arrive to the same 
conclusions by considering that the dashed lines originate from the exchange of an 
arbitrary heavy meson (the only relevant point in our derivations is the bilinear 
character of the meson-$\bar{N}N$ interaction in the nucleon fields). In this way 
it is straightforward 
to realize the presence of a factor $1/2$ in front of fig.3d together with the 
factors $\prod_j (-1)^{n_j+1}/n_j$ which here is simply $1$. 
The relation between contact four nucleon interactions at low energies in effective 
field theories and its saturation by integrating-out heavy meson resonances has been 
established in ref.\cite{epe2}.

\begin{figure}[htb]
\centerline{\epsfig{file=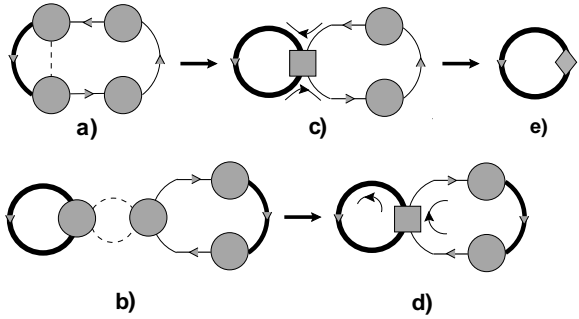,width=3.0in}}
\vspace{0.3cm}

\caption[pilf]{\protect \small
Diagrams 3a and 3b represent some typical many-particle diagrams generated from 
eqs.(\ref{fd}), (\ref{fZ}). The circle indicates an $A$ operator insertion (which 
in addition can have  attached to it more lines than shown) and the dashed line 
corresponds to a pion exchange. Figs.3c and 3d arise by considering  the 
ultra-vioalet divergent part of the pion loops leading to local terms denoted by 
squares. Fig.3e is a local $\pi N$ counterterm, indicated by a diamond. For more details 
see the text.}
\end{figure}

There is still an important difference to be discussed when comparing figs.3c and 3d, 
which in fact is related to the presence of the factor (det$\,D$) in all the 
formulae from eq.(\ref{tran8}) to eq.(\ref{fZ}). The latter corresponds to contributions
 to the chiral Lagragian from closed fermion loops in the vacuum and in the 
spirit of the effective field theories these contributions, coming from states with 
masses close to or above the chiral scale $\Lambda_\chi\approx M_\rho$, are 
incorporated in the counterterms of the vacuum effectrive field theory.
 In this way we will set det$\,D=1$ in the following. 
Indeed on the right hand side of the square of fig.3c we can recognize a momentum 
loop flowing along the free baryon propagators without any insertion of a Fermi-sea. 
This simply means that since all the momenta that could go into this closed momentum 
loop are soft because they come from the circles, it just  
corresponds to vacuum renormalization from heavy particles and is reabsorbed 
in the low energy counterterms. This is schematically shown in fig.3e with a diamond  
corresponding to a higher order $\pi N$ counterterm. 
This is consistent as far as one is restricted 
to the low energy and momentum regime.  However, when a Fermi-sea baryon line is 
present the momentum running though 
the baryon lines is of the form $p_j=p+Q_j$ with $p=(E(p),\bold{p})$ and 
$|{\bold{p}}|<k_F$. Thus one is inserting a medium parameter $k_F$ and shifting 
upward up to around $m_N$ the energy level around which one is perturbating. 

\vskip 15pt
\noindent
{\bf 5. }We now turn to the generalization of the formalism to the case of asymmetric nuclear 
matter,  with different densities of neutrons, $\rho_n$, and protons, 
$\rho_p$, with Fermi momenta $k_F^{(n)}=(3\pi^2 \rho_n)^{1/3}$ and 
$k_F^{(p)}=(3\pi^2 \rho_p)^{1/3}$, respectively. Following the previous derivation of 
eq.(\ref{fd}) one can easily convince oneself that the only change is to remove 
the sum over 
isospin indices  and to distinguish between $\alpha=1$ (proton) and $\alpha=2$ 
(neutron). In this way we will have:
\ba
\label{fdg}
&&e^{i \cZ[v,a,s,p]}=\!\int\! [dU] \exp\Bigg\{i\!\int\! dx \cL_{\pi\pi} 
+\sum_{r=1}^2
 \nonumber \\ &&  
\!\int^{k_F^{(p)}}\hspace{-0.5cm} \frac{d\vp}{(2\pi)^3 2E(p)_1} \!\int\! d\vx \!\int\! 
d\vy 
e^{-i\vp \vx} u^\dagger_{r\,1}(\vp)\,\log[\cF]
\arrowvert_{(\vx\,1,\vy\,1)} \nonumber \\ && 
e^{i\vp \vy} u_{r\,1}(\vp)+\sum_{r=1}^2 \!\int^{k_F^{(n)}}\hspace{-0.5cm} 
\frac{d\vp}{(2\pi)^3 2E(p)_2}\!\int\! d\vx \!\int\! d\vy\, 
e^{-i\vp \vx} 
\nonumber \\ &&
u^\dagger_{r\,2}(\vp)\,\log[\cF]
\arrowvert_{(\vx\,2,\vy\,2)}\,e^{i\vp \vy} u_{r\,2}(\vp)\Bigg\}
\ea
Notice that we have indicated separately the energies of protons $E(p)_1$ and 
neutrons $E(p)_2$ with three-momentum $\vp$, since the previous equation is valid for 
the non-isospin limit as well. Nevertheless, in order to simplify the formulae, we will 
consider in the following the case with equal nucleon masses. We also introduce the 
$2\times 2$ matrix: 
\ba
\label{as}
n(p)&=&\left( \begin{array}{cc}\theta(k_F^{(p)}-p) & 0\\ 0 & \theta(k_F^{(n)}-p) 
\end{array}\right)
\nonumber \\
&\equiv&\left( \begin{array}{cc}n(p)_1 & 0\\ 0 & n(p)_2 \end{array}
\right) \nonumber \\
&=&\hat{n}(p)I_2+\bar{n}(p)\tau_3
\ea
with $I_2$ the $2\times 2$ unity matrix, $\tau_3$ the usual Pauli matrix 
$diag(1,-1)$,  $\hat{n}(p)=(n(p)_1+n(p)_2)/2$ and $\bar{n}(p)=
(n(p)_1-n(p)_2)/2$. Then eq.(\ref{fdg}) can be rewritten as:

\ba
\label{fZ2}
&&e^{i \cZ[v,a,s,p]}=\!\int\! [dU] \exp\bigg\{i\!\int\! dx \cL_{\pi\pi} \!\!
+\!\!\int\!\! d\vx \!\!\int\!\! d\vy\!\!\int\hspace{-0.2cm} \frac{d\vp}{(2\pi)^3 2E(p)}
\nonumber \\ && e^{-i\vp (\vx-\vy)} \hbox{Tr}\bigg( 
n(p) \log[\cF]\arrowvert_{(\vx,\vy)} (\barr{p}+m)\bigg)\bigg\}=
\nonumber \\&&
=\!\int\! [dU] \exp\Bigg\{i\!\int\! dx\, \cL_{\pi\pi}-i
\!\int\!\frac{d\vp}{(2\pi)^3 2 E(p)} \!\int\! dx\,dy\, e^{ip(x-y)}\,
\nonumber \\&&
\hbox{Tr}\Bigg( A[I_4-D_0^{-1}A]^{-1}
\arrowvert_{(x,y)}
(\barr{p+m_N})\,n(p)\Bigg)+
\nonumber \\ &&
\frac{1}{2}\!\int\!\!\! \frac{d\vp}{(2\pi)^3 2E(p)}\!\int\! \!\! 
\frac{d\vq}{(2\pi)^3 2 E(q)}
\!\int\! dx\,dx'\,dy\,dy'\,e^{ip(x-y)}
\nonumber \\ &&e^{-iq(x'-y')}\,
\hbox{Tr}\Bigg(
 A[I_4-D_0^{-1}A]^{-1}\arrowvert_{(x,x')}\, (\barr{q}+m_N)\,n(q)
\nonumber \\ &&
A[I_4-D_0^{-1}A]^{-1}\arrowvert_{(y',y)}(\barr{p}+m_N)\,n(p)\Bigg)+...\Bigg\}.
\ea
Comparing this equation with eq.(\ref{fZ}), the only difference in the 
Feynman rules is the inclusion of an isospin 
matrix $n(p)$ associated to every Fermi-sea insertion.

The case of a polarized nuclear matter can be treated in the same way just by 
doing in eq.(\ref{fdg}) the replacement: 
\ba
\label{unpola}
&&\sum_{r=1}^2\!\int^{k_F}\hspace{-0.1cm} d\vp\, u^\dagger_{r \alpha}(\vp)\;...
\;u_{r\alpha}(\vp) \rightarrow 
\!\int^{k_F}\hspace{-0.1cm} d\vp\, u^\dagger_{r_1 \alpha}(\vp)\;...\;u_{r_1 \alpha}(\vp)
\nonumber \\&&
+\!\int^{k_F}\hspace{-0.1cm} d\vp\, u^\dagger_{r_2 \alpha}(\vp)\;...\;u_{r_2 \alpha}(\vp)
\ea
since there are two spin states per momentum state.

\vskip 15pt
\noindent
{\bf 6. }The present formalism is applied in ref.\cite{med2} to evaluating several quantities 
relevant for low energy QCD in the nuclear medium. There we will also address 
in detail the issue of the chiral counting from eq.(\ref{fZ2}) in the nuclear 
medium and the limitations of a plain perturbative treatment of CHPT at finite density. 
Just to mention that instead of a relativistic treatment of the baryons we could 
also have considered in the same way the non-relativistic case. This makes more 
straightforward the evaluation of baryon loops \cite{fur} although one has also to 
take into account recent developments in the field of effective field theories 
with propagation of relativistic heavy particles \cite{leut,goit,leh}. In any case 
the main issue still to be addressed in the medium, as discussed in refs.
\cite{fur,med2}, is to deal with the problem of the large S-wave scattering lengths 
in the nucleon-nucleon scattering which introduces a new extra scale of  only $\sim$10 MeV 
alredy in the vaccum case, where consistent power counting schemes have been 
developed\cite{kap}. 
Some interesting findings in this direction, requiring further consideration, can be 
found in ref.\cite{steele} for a theory without pions.

To conclude, we have derived the $SU(2)$ chiral Lagrangian with extenal 
sources in 
the presence of non-zero nuclear density by explicitely working out in quantum field 
theory the in-medium contributions from the $\pi N$ chiral 
Lagrangian, eq.(\ref{fZ})(symmetric nuclear matter), eq.(\ref{fZ2}) 
(asymmetric nuclear matter) and 
eq.(\ref{unpola}) (asymmetric and unpolarized nuclear matter). Then the perturbative 
theory is developed and the corresponding Feynman rules are given. Contrarily to 
standard many-body techniques, the rules and diagrams here derived for the general 
relativistc case are analogous to the usual ones from 
vacuum quantum field theory without modification of the baryon propagators establishing 
a neat separation between in-medium and vacuum contributions, as indicated in figs.1 
and 2. Applications of this many-particle relativistic quantum field 
theory formalism to actual calculations can be found in ref.\cite{med2}.

I would like to thank Andreas~Wirzba and Ulf-G.~Mei{\ss}ner for useful discussions and 
for a critical reading of the manuscript. This work was supported in part by funds from
DGICYT under contract PB96-0753 and from the EU TMR network Eurodaphne, contract
no. ERBFMRX-CT98-0169.


\end{document}